%% file: DS_codes_v0.73-1.tex
\documentclass[conference,letterpaper]{IEEEtran}

\usepackage[utf8]{inputenc} 
\usepackage[T1]{fontenc}
\usepackage{url}
\usepackage{ifthen}
\usepackage{cite}
\usepackage[cmex10]{amsmath} % Use the [cmex10] option to ensure complicance
\usepackage{amssymb}
\usepackage{algorithm}
\usepackage{graphicx}
\usepackage{xcolor}
\usepackage{tikz}
\usepackage{braket}

\usepackage[caption=false,labelformat=simple]{subfig}		% the latter command to show Fig.~1(a) instead of Fig.~1a

\newcommand{\dlt}{\delta}

\newcommand{\sC}{{\cal C}}
\newcommand{\sG}{{\cal G}}
\newcommand{\sM}{{\cal M}}
\newcommand{\sN}{{\cal N}}
\newcommand{\sS}{{\cal S}}

\DeclareMathOperator*{\argmax}{arg\,max}

\usepackage{amsthm}
\newtheorem{definition}{Definition}

\newtheorem{theorem}{Theorem}

\DeclareMathOperator{\wt}{wt}

\usepackage[colorinlistoftodos,prependcaption]{todonotes}
\usepackage{xcolor}
\usepackage{xargs}

\newcommandx{\yellownote}[2][1=]{\todo[inline,linecolor=yellow,backgroundcolor=yellow!25,bordercolor=yellow,#1]{#2}}

\begin{document}
%\title{On the Decoding of Quantum Data-Syndrome Codes} 
%\title{A Belief-Propagation Decoding Algorithm for the Quantum Data-Syndrome Codes} 
\title{Decoding of Quantum Data-Syndrome Codes via Belief Propagation} 

% %%% Single author, or several authors with same affiliation:
%\author{%
%  \IEEEauthorblockN{Kao-Yueh Kuo}
%  \IEEEauthorblockA{Institute of Communications Engineering\\
%			National Chiao Tung University, Hsinchu 30010, Taiwan\\
%                    Email:  kykuo@nctu.edu.tw}
%  \and
%  \IEEEauthorblockN{Ching-Yi Lai}
%  \IEEEauthorblockA{Institute of Communications Engineering\\
%			National Chiao Tung University, Hsinchu 30010, Taiwan\\
%                    Email: cylai@nctu.edu.tw}
%}

%%% Several authors with up to three affiliations:
 \author{%
   \IEEEauthorblockN{Kao-Yueh Kuo\IEEEauthorrefmark{1}\IEEEauthorrefmark{3},
                     I-Chun Chern\IEEEauthorrefmark{2}\IEEEauthorrefmark{4},
                     and Ching-Yi Lai\IEEEauthorrefmark{1}\IEEEauthorrefmark{2}\IEEEauthorrefmark{3}\IEEEauthorrefmark{4}}
%   \IEEEauthorblockA{\IEEEauthorrefmark{1}%
%                     Department of Electrical and Computer Engineering,\\
%                     National Chiao Tung University, Hsinchu 30010, Taiwan,
%                     ethanc.eed06@nctu.edu.tw}
%   \IEEEauthorblockA{\IEEEauthorrefmark{2}%
%                     Institute of Communications Engineering,\\
%                     National Chiao Tung University, Hsinchu 30010, Taiwan,
%                     \{kykuo, cylai\}@nctu.edu.tw}
   \IEEEauthorblockA{\IEEEauthorrefmark{1}%
                     Institute of Communications Engineering and
                     \IEEEauthorrefmark{2}%
                     Department of Electrical and Computer Engineering,\\
                     National Yang Ming Chiao Tung University, Hsinchu 30010, Taiwan\\
                     \IEEEauthorrefmark{3}%
                     Institute of Communications Engineering and
                     \IEEEauthorrefmark{4}%
                     Department of Electrical and Computer Engineering,\\
                     National Chiao Tung University, Hsinchu 30010, Taiwan,
                     \{kykuo, ethanc.eed06, cylai\}@nctu.edu.tw}
 }

%%% Many authors with many affiliations:
% \author{%
%   \IEEEauthorblockN{Albus Dumbledore\IEEEauthorrefmark{1},
%                     Olympe Maxime\IEEEauthorrefmark{2},
%                     Stefan M.~Moser\IEEEauthorrefmark{3}\IEEEauthorrefmark{4},
%                     and Harry Potter\IEEEauthorrefmark{1}}
%   \IEEEauthorblockA{\IEEEauthorrefmark{1}%
%                     Hogwarts School of Witchcraft and Wizardry,
%                     1714 Hogsmeade, Scotland,
%                     \{dumbledore, potter\}@hogwarts.edu}
%   \IEEEauthorblockA{\IEEEauthorrefmark{2}%
%                     Beauxbatons Academy of Magic,
%                     1290 Pyrénées, France,
%                     maxime@beauxbatons.edu}
%   \IEEEauthorblockA{\IEEEauthorrefmark{3}%
%                     ETH Zürich, ISI (D-ITET), ETH Zentrum, 
%                     CH-8092 Zürich, Switzerland,
%                     moser@isi.ee.ethz.ch}
%   \IEEEauthorblockA{\IEEEauthorrefmark{4}%
%                     National Chiao Tung University (NCTU), 
%                     Hsinchu, Taiwan,
%                     moser@isi.ee.ethz.ch}
% }

\maketitle

%%%%%%
%% Abstract: 
%% If your paper is eligible for the student paper award, please add
%% the comment "THIS PAPER IS ELIGIBLE FOR THE STUDENT PAPER
%% AWARD." as a first line in the abstract. 
%% For the final version of the accepted paper, please do not forget
%% to remove this comment!
%%
\begin{abstract}
Quantum error correction is necessary to protect logical quantum states and operations. However, no meaningful data protection can be made when the syndrome extraction is erroneous due to faulty measurement gates. Quantum data-syndrome (DS) codes are designed to protect the data qubits and syndrome bits concurrently. In this paper,  we  propose an efficient decoding algorithm for quantum DS codes with sparse check matrices. Based on a refined belief propagation (BP) decoding for stabilizer codes, we propose a DS-BP algorithm to handle the quaternary quantum data errors and binary syndrome bit errors. Moreover, a sparse quantum code may  inherently be able to handle minor syndrome errors so that fewer redundant syndrome measurements are necessary. We demonstrate this with simulations on a quantum hypergraph-product code. 
\end{abstract}

\section{Introduction}
Quantum error-correction codes  are  indispensable in fault-tolerant quantum computation (FTQC) and quantum communication \cite{Sho96,KL97,Ste07,LB17}. 
Quantum stabilizer codes are an important class of quantum codes \cite{GotPhD,CRSS98,NC00} since they allow simple encoding and decoding processes  similar to classical error-correcting codes.  This decoding process is analogous to the classical syndrome-based decoding. 
However, quantum operations are inevitably faulty, making robust syndrome extraction difficult.
Using imperfect syndrome measurements, the outcome will be inaccurate and it may also corrupt the data qubits. 
As a consequence, we will be led into using a wrong syndrome, which will in turn provide us with an incorrect decoding result for error recovery.

Conventionally,  an error recovery operation is chosen by repeated syndrome measurements, followed by a certain decision strategy~\cite{Sho96}. 
For topological codes, the minimum-weight perfect-matching decoder is able to locate a likely error with high complexity~\cite{Fow+12}.  
For the ease of analysis,  a simple error model is considered, where each qubit independently suffers a Pauli error and each syndrome bit independently suffers a bit-flip error. 
It has been shown that   quantum stabilizer codes could be capable of correcting data  errors and syndrome errors simultaneously,
known as \textit{quantum data-syndrome (DS) codes}~\cite{ALB14,Fuji14,ALB16,ALB20}.
Recently, there has been a proposal put forward to construct quantum convolutional DS codes with a generalized Viterbi decoding algorithm~\cite{ZAWP19}.
Following Bombin's seminal work on the single-shot fault-tolerant quantum error-correction to handle measurement errors, other approaches  have also been proposed for topological codes \cite{Bom15,BNB16,BDMT16}.

We would like to develop a decoding algorithm with low complexity for FTQC by first studying the simple error model.
In this paper we study the decoding of sparse-graph quantum codes~\cite{Kit03,MMM04,TZ14} with independent data and syndrome errors 
using belief propagation (BP). BP has been shown to be powerful and efficient for  classical decoding \cite{Gal63,Mac99},  artificial intelligence decision \cite{Pea88}, and  quantum decoding \cite{MMM04,PC08,KL20}.
Since the Pauli errors $\{I,X,Y,Z\}$ are quaternary but a syndrome bit-flip error is binary, a hybrid quaternary-binary nature of the DS codes is created,
which makes their decoding complicated. 
Previously BP for quantum codes  (BP$_4$) has been refined to pass only scalar messages because of the fact that the error syndromes for a stabilizer code are binary~\cite{KL20}.
Accordingly, we propose a decoding algorithm (called \mbox{\emph{DS-BP$_4$}}) for quantum DS codes, showing that
the quaternary data error information and binary syndrome \mbox{bit-flip} information can be handled by scalar message passing.

A quantum DS code has additional redundant stabilizers being measured and usually a two-stage sequential decoding process is applied~\cite{ALB14,ZAWP19}.
In reality, syndrome measurements take a longer time than simple logical operations.
One would like to have redundant measurements as few as possible.
We demonstrate that DS-BP$_4$ on a  quantum hypergraph-product (HP) code without any redundant syndrome bits  may have performance close to the case of perfect syndrome measurements (where the loss in block error rate is less than an order for the $[[129,28]]$ HP code).
On the other hand, we assume that the error rate is higher if repeated syndrome measurements are conducted since this usually takes a longer time. Then DS-BP$_4$ on the HP code without  any redundant syndrome bits performs better
than the two-stage sequential decoding.

The paper is organized as follows. In Sec.~\ref{sec:Stb}, we introduce the basics of stabilizer codes and quantum DS codes. 
In Sec.~\ref{sec:Alg},  a DS-BP$_4$ decoding algorithm for quantum DS codes is proposed along with the simulation results.
Then we conclude in Sec.~\ref{sec:Conc}. %,  and discuss some future work.

\section{Quantum Data-Syndrome Codes} \label{sec:Stb}
\subsection{Quantum Stabilizer Codes} 
We consider binary stabilizer codes \cite{GotPhD,NC00}.
Suppose that $\sS$ is a subgroup of the $N$-fold Pauli group $\sG_N$ and  generated by $N-K$ independent $N$-fold Pauli operators $S_1,S_2,\dots,S_{N-K}$ such that $S_mS_{m'}=S_{m'}S_m$ and $-I^{\otimes N} \notin \sS$. 
An $[[N,K]]$ stabilizer code $C(\sS)$ encodes $K$ logical qubits into $N$ physical qubits
and its code space is  the joint $(+1)$ eigenspace of the elements in $\sS$.
The elements in $\sS$ are called \emph{stabilizers}.
If a Pauli error anticommutes with some stabilizers, it can be detected by measuring the eigenvalues of the stabilizers. 
Thus the measurement outcomes, called  \textit{error syndrome}, are used to determine a correction operation.
Sometimes additional    redundant stabilizers $\{S_m\}_{m=N-K+1}^M$ are measured
for enhancing BP in decoding \cite{ROJ19} or  handling syndrome errors  \cite{ALB16,ALB20}.

Without loss of generality, $S_m$ is of the form $S_m =   S_{m1} \otimes S_{m2} \otimes \dots \otimes S_{mN},$ 
where   $S_{mn} \in \{I=[{1\atop 0}{0\atop 1}], X=[{0\atop 1}{1\atop 0}], Z=[{1\atop 0}{0\atop -1}], Y=iXZ\}$ for $n=1,2,\dots,N$.
We will ignore the notation $\otimes$ without confusion.
Then 
	$$S = [S_{mn}]\in\{I,X,Y,Z\}^{M\times N}$$
	is called the   \emph{check matrix} of the stabilizer code. 
Two Pauli operators of the same dimension, $E$ and $F$,  either commute or anticommute with each other. We define 
\begin{align}
\langle E,F\rangle_{\sG} = 
\begin{cases}
 0, & \mbox{if  $EF=FE$};\\
 1, & \mbox{otherwise}.
\end{cases}
\end{align}
For an  error  $E=E_1E_2\cdots E_N\in\sG_N$, its \textit{binary error syndrome}  is given by
 $z=(z_1,z_2,\dots,z_M)\in\{0,1\}^M$, where
$$ z_m = \langle E, S_m\rangle_{\sG} = \sum_{n=1}^N \langle E_n, S_{mn}\rangle_{\sG} \mod 2.$$
Let $\wt_\sG(E)$ denote the number of non-identity entries in $E\in\sG_N$.
The \emph{minimum distance}  of $C(\sS)$ is defined as
$$d = \min\{\wt_\sG(E) \mid E\in \{I,X,Y,Z\}^N\setminus\sS,\, \langle E,S_m \rangle = 0 ~\forall~ m\}.$$
A stabilizer code with minimum distance $d$ can correct any errors $E$ with $\wt_\sG(E)\le t$, where $t = \lfloor\frac{d-1}{2}\rfloor$.\\

\subsection{Quantum Data-Syndrome (DS) Codes}

In addition to a Pauli error $E\in\sG_N$ on the data qubits,  each syndrome bit $z_m$ suffers an independent bit-flip error $e_m\in\{0,1\}$.
Now the syndrome bit relation becomes 
\begin{equation} \label{eq:nos_z_m}
z_m =  \langle E,S_m \rangle_{\sG} + e_m \mod 2.
\end{equation}
Consequently the  check matrix for the DS code is defined by
\begin{align}
\tilde{S}  = [S~I_M],
\end{align}
where $I_M$ is an $M\times M$ binary identity matrix.
The $m$-th row of $\tilde{S}$ is denoted by $\tilde{S}_m = (S_m,(I_M)_m) ~\in~ \{I,X,Y,Z\}^N\times\{0,1\}^M$.
Define the product of   $e,f\in\{0,1\}^M$ by 
$\langle e,f \rangle_b = \sum_{j=1}^M e_jf_j \mod 2.$
Then the  product of $(E,e),(F,f)\in \sG_N\times \{0,1\}^M$ is defined by 
\begin{align}\langle(E,e),(F,f)\rangle= \langle E,F\rangle_{\sG}+\langle e,f \rangle_b   \mod 2.
\end{align}
For $e\in\{0,1\}^M$, let
$\wt_b(e)$ denote the number of its nonzero entries.
Then the weight of $(E,e)$ is defined as  $$\wt(E,e) = \wt_\sG(E)+\wt_b(e).$$

\begin{definition} \label{def:DS}
	Let $S\in\{I,X,Y,Z\}^{M\times N}$ be a check matrix of an $[[N,K]]$ stabilizer code, where $M\ge N-K$. %and there are $M-(N-K)$ redundant measurements.
	We say that $S$ induces an $[[N,K\,|\,M]]$ quantum DS code $\tilde{\sC}$ with a DS check matrix $\tilde{S} = [S~I_M]$, where
	$$\tilde{\sC}= \{ (F,f) \in \{I,X,Y,Z\}^N\times\{0,1\}^M \mid \langle (F,f),\tilde{S}_m \rangle = 0 ~\forall~ m \}.$$
	The  \emph{(DS) minimum distance} of  $\tilde{\sC}$ is defined as
	$$\tilde{d} = \min\{\wt(F,f) \mid (F,f)\in\tilde{\sC}\setminus \tilde{\sS}\},$$
	where $\tilde{\sS} \triangleq \{(F,\pmb 0) \mid F\in\sS\}$.
	\normalsize
\end{definition}

\begin{theorem}\cite{ALB20} 
	An $[[N,K\,|\,M]]$ quantum DS code with DS minimum distance $\tilde{d}$ can correct any error $(E,e)\in\sG_N\times\{0,1\}^M$ with $\wt(E,e)\le \tilde{t}$, where $ \tilde{t} = \lfloor\frac{ \tilde{d}-1}{2}\rfloor$. 
\end{theorem}

\section{Belief Propagation for Quantum DS Codes} \label{sec:Alg}

The minimum distance $d$ of a stabilizer code is an upper bound on its induced DS minimum distance $\tilde d$. 
The DS minimum distance usually achieves this upper bound by additional redundant measurements \cite{ALB16,ALB20,ZAWP19}.

In some conditions, we can have the syndrome protected without redundant measurements~\cite{Fuji14} and this is a desired property especially when the syndrome measurements are expensive: the syndrome error rate could be higher than the data error rate since a syndrome measurement involves many two-qubit operations and single qubit measurements; also more syndrome measurements take more processing time, which in turn incurs a higher data error rate. 

In the next section, we will simulate a case of CSS codes~\cite{CS96,Ste96} since they are commonly used in FTQC. 
The stabilizer group of a CSS code can be chosen to be products of only $X$ operators or only $Z$ operators.
Thus we have a binary check matrix
$\textstyle   H = [{H_X\atop O}\mid{O\atop H_Z}],$ 
where $H_X\in\{0,1\}^{M_1\times N}$, $H_Z\in\{0,1\}^{M_2\times N}$, $M=M_1+M_2$, and $H_X H_Z^T = O$, where $O$ is the all-zero matrix whose dimension can be inferred from the context.
The induced quantum DS code has a (binary) check matrix:
\begin{equation} \label{eq:CSS_DS_H}
\tilde{H} = \textstyle [{H_X\atop O}{I_{M_1}\atop O}\mid{O\atop H_Z}{O\atop I_{M_2}}].
\end{equation}
It is not hard to obtain the following theorem.
\begin{theorem}\label{thm:c2d3}
	Let $\tilde{H}$ be defined as in (\ref{eq:CSS_DS_H}), where $H_X$ and $H_Z$ are the parity-check matrices of two classical codes, respectively, each with minimum distance  at least~3. 
	If  every column of  $H_X$~and $H_Z$  is of weight at least~2, then the induced quantum DS code  has   minimum distance at least~3.
\end{theorem}

The conditions can be generalized for higher-weight errors but it is complicated. We remark that the minimum distance of a code provides only a reference for its error performance; we care more about the following practical decoding problem:

\noindent{\bf The quantum DS decoding problem}: Given a DS check matrix $\tilde{S} = [S~I_M]$, where $S\in\{I,X,Y,Z\}^{M\times N}$, a binary syndrome $z\in\{0,1\}^M$ of  $(E,e)\in\sG_N\times \{0,1\}^M$, and certain characteristics of the error model, the decoder has to infer   $(\hat E, \hat e)$, where $\hat E\in\{I,X,Y,Z\}^N$ and $\hat e\in\{0,1\}^M$ such that $\langle (\hat E,\hat e), \tilde{S}_m \rangle = z_m$ for all $m=1,2,\dots,M$ and $\hat E\in E\sS$.

\subsection{The DS-BP Algorithm}

	We would like to have a successful decoding with  probability   as high as possible (according to a specific error model).
		In general, achieving an optimum decoding is  extremely difficult for conventional stabilizer codes \cite{KL13_20,IP15} and this is also the case for DS codes.
			However, if $S$ is a sparse matrix,  then $\tilde{S}$ is also sparse. Using BP to decode DS codes is also efficient and possible to have good performance like stabilizer codes \cite{MMM04,PC08,KL20}.

\begin{figure}[hbtp] \centering  \resizebox{0.4\textwidth}{!}{~~ \input{Fig_3bits_DS}}
	\caption{The Tanner graph of $\tilde{S}= [{X\atop Z}{Y\atop Z}{I\atop Y}{1\atop 0}{0\atop 1}]$. 
		There are three types of edges connecting a data node and a check node.
	} \label{fig:hatS}
\end{figure}
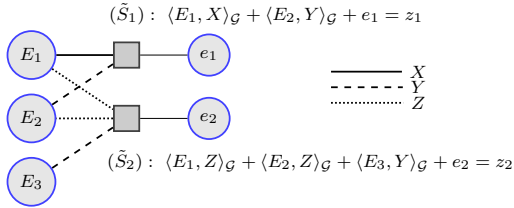

The DS check matrix $\tilde{S}=[S~I_M]$ corresponds to a Tanner graph consisting of $N$ data(-variable) nodes, $M$ \mbox{syndrome(-variable)} nodes, and $M$ check nodes. There is an edge connecting data node $n$ and check node $m$ if $S_{mn}\ne I$. There is always an edge connecting check node $m$ and syndrome node~$m$.
For example, if  $S=[{X\atop Z}{Y\atop Z}{I\atop Y}]$, then $\tilde{S} = [S~I_M] = [{X\atop Z}{Y\atop Z}{I\atop Y}{1\atop 0}{0\atop 1}]$ and its corresponding Tanner graph is shown in Fig.~\ref{fig:hatS}.

Suppose that the data nodes and syndrome nodes are numbered from $1$ to $N+M$. 
Let $\sM(n)$ be the neighboring check nodes   of a variable node $n$,
and $\sN(m)$ be the neighboring variable nodes   of a check node $m$.
Using a similar derivation in \cite[Algorithm~3]{KL20} (or more simply, in~\cite{KL20b}), it is not so difficult to generalize the quaternary BP (BP$_4$) algorithm  for stabilizer codes \cite[Algorithm~3]{KL20} to DS-BP$_4$ for  quantum DS codes as in Algorithm~\ref{alg:QBP} that handles only scalar messages.

The initial probabilities of data and syndrome errors can be assigned in DS-BP$_4$. 
Suppose that each qubit suffers a memoryless depolarizing channel  of depolarizing rate   $\epsilon_D$
and each syndrome bit suffers a memoryless binary symmetric channel (BSC) of crossover rate  $\epsilon_S$.
Then  $(p_n^I, p_n^X, p_n^Y, p_n^Z)$  are initialized to $(1-\epsilon_D, \frac{\epsilon_D}{3}, \frac{\epsilon_D}{3}, \frac{\epsilon_D}{3})$ for each data-variable node $n\in\{1,\dots,N\}$, and   $(p_n^{(0)},p_n^{(1)}) $ are initialized to $(1-\epsilon_S, \epsilon_S)$ for each syndrome-variable node $n\in\{N+1,\dots,N+M\}$.

\begin{algorithm}[ht!]
	\caption{: Quaternary DS-BP decoding for quantum DS codes with a parallel schedule (parallel DS-BP$_4$)} \label{alg:QBP}
	\textbf{Input}: $\tilde{S}=[S\ I_{M}]$, ${S}\in\{I,X,Y,Z\}^{M\times N}$,\, $z \in\{0,1\}^M$, and initial $\{(p_n^I, p_n^X, p_n^Y, p_n^Z)\}_{n=1}^N, \{(p_n^{(1)}, p_n^{(0)})\}_{n=N+1}^{N+M}$.\\
	{\bf Initialization.}  
	For all $n=1$ to $N+M$ and all $m\in\sM(n)$:
	\begin{itemize} 
		\item If $n\leq N$, let $q_{mn}^W = p_n^W$ for $W\in\{I,X,Y,Z\}$,\\
		and let $q_{mn}^{(0)} = q_{mn}^I+q_{mn}^{S_{mn}}$ and $q_{mn}^{(1)} = 1 - q_{mn}^{(0)}$.
		\item If $n> N$, let $q_{mn}^{(0)} = p_n^{(0)}$ and $q_{mn}^{(1)} = p_n^{(1)}$.
		\item Calculate
		\begin{equation}
		d_{n\to m}=q_{mn}^{(0)}-q_{mn}^{(1)}. \label{eq:dmn_BP4}
		\end{equation}
	\end{itemize}
	{\bf Horizontal Step.} 
	For all $m=1$ to $M$ and all $n\in\sN(m)$:
	\begin{itemize}
		\item Compute
		\begin{align}
		\dlt_{m\to n} = (-1)^{z_m}\prod_{n'\in\sN(m)\setminus n} d_{mn'}.   \label{eq:delta_mn_BP4}
		\end{align}
	\end{itemize}
	{\bf Vertical Step.} 
	For all $n=1$ to $N+M$ and all $m\in\sM(n)$: 
	\begin{itemize}
		\item Let $r_{mn}^{(0)} = (1+\dlt_{mn})/2$ and $r_{mn}^{(1)} = (1-\dlt_{mn})/2$.
		\item If $n\le N$, compute
		\begin{align}
		q_{mn}^W &= p_n^W\prod_{m'\in\sM(n)\setminus m} r_{m'n}^{(\langle W, S_{m'n}\rangle)}, ~~  W\in\{I,X,Y,Z\}, \label{eq:qmnW}\\
		q_{mn}^{(0)} &= a_{mn}(q_{mn}^I + q_{mn}^{S_{mn}}), \notag\\
		q_{mn}^{(1)} &= a_{mn}(\textstyle \sum_{W'\in\{X,Y,Z\}\setminus S_{mn}}q_{mn}^{W'}). \notag
		\end{align}
		\item If $n> N$, compute
		\begin{align}
		q_{mn}^{(b)} &= a_{mn}\,p_n^{(b)}\prod_{m'\in\sM(n)\setminus m} r_{m'n}^{(b)}, ~~ b\in\{0,1\}. \label{eq:qmn1}
		\end{align}
		\item Each $a_{mn}$ is a chosen scalar such that $q_{mn}^{(0)}+q_{mn}^{(1)}=1$.
		\item Update: $d_{n\to m} = q_{mn}^{(0)} - q_{mn}^{(1)}$. 
	\end{itemize}
	{\bf Hard Decision.} 
	For all $n=1$ to $N+M$:
	\begin{itemize}
		\item If $n\le N$, compute
		\begin{align*}
		q_n^W &= p_n^W\prod_{m\in\sM(n)} r_{mn}^{(\langle W,S_{mn}\rangle)}, ~~ W\in\{I,X,Y,Z\}, %\label{eq:qnW}
		\end{align*} 
		and let $\hat E_n = \argmax_{W\in\{I,X,Y,Z\}} q_n^{W}.$
		\item If $n> N$, compute
		\begin{align*}
		q_n^{(b)} &= p_n^{(b)}\prod_{m\in\sM(n)} r_{mn}^{(b)}, ~~ b\in\{0,1\}, %\label{eq:qn1}
		\end{align*}
		and let $\hat e_n = 0$, if $q_n^{(0)}>q_n^{(1)}$, and $\hat e_n = 1$, otherwise.
		\item \mbox{Let $\hat E = \hat E_1\hat E_2\cdots\hat E_N$ and $\hat{e} = (\hat e_{N+1},\hat e_{N+2},\dots,\hat e_{N+M})$}.
		\begin{itemize}
			\item If $\langle (\hat E, \hat e), \tilde{S}_{m} \rangle = z_m$ for all $m=1$ to $M$, halt and return ``CONVERGED''.
			\item Otherwise, if a maximum number of iterations is reached, halt   and return ``FAIL''.
			\item Otherwise, repeat from the horizontal step.
		\end{itemize}
	\end{itemize}
\end{algorithm}

Algorithm~\ref{alg:QBP} performs the update according to a parallel schedule \cite{KL20} and is referred to as \emph{parallel DS-BP$_4$}.
We also consider an update order according to a serial schedule running along the check nodes as in Algorithm~\ref{alg:SBP}, which is referred to as \emph{serial DS-BP$_4$}. 
Consider  the example in Fig.~\ref{fig:hatS} again. Its message update order using parallel DS-BP$_4$ (resp. \mbox{serial DS-BP$_4$}) is illustrated in Fig.~\ref{fig:PBP} (resp. Fig.~\ref{fig:SBP}).
The message update order plays an important role in BP, especially when the Tanner graph has numerous short cycles~\cite{KL20,RV20}.

\begin{algorithm}[t] 
	\caption{: Quaternary DS-BP decoding  with a serial schedule along the check nodes (serial DS-BP$_4$)}\label{alg:SBP}
	\textbf{Input}: The same as in Algorithm~\ref{alg:QBP}.\\
	{\bf Initialization.} 
	For all $m=1$ to $M$ and all $n\in\sN(m)$:
	\begin{itemize}
		\item Let $\delta_{m\to n}=0$.
	\end{itemize}
	{\bf Serial Update.} For each check node $m=1$ to $M$:%\\
	\begin{itemize}
		\item For every $n\in\sN(m)$, do the same as in the five bullet points around \eqref{eq:qmnW} and \eqref{eq:qmn1}, with the order specified here.
		\item For every $n\in\sN(m)$, compute
		\begin{align}
		\dlt_{m\to n} = (-1)^{z_m}\prod_{n'\in\sN(m)\setminus n} d_{mn'}.   \label{eq:delta_mn_BP4}
		\end{align}
	\end{itemize}
	
	{\bf Hard Decision.} 
	\begin{itemize}
		\item Do  as in Algorithm~\ref{alg:QBP},   except that ``repeat from
		the horizontal step" must be replaced by ``repeat from the serial
		update step".
	\end{itemize}
	
\end{algorithm}

\begin{figure}[h!]	   % use \resizebox{rowsize}{colsize} and {!} means to maintain the aspect ratio
	\centering
	\resizebox{0.3\textwidth}{!}{ 
		\subfloat[\label{fig:PBP_1}]{\input{Fig_PBP_1.tex}}\qquad\qquad
		\subfloat[\label{fig:PBP_2}]{\input{Fig_PBP_2.tex}}
	}
	\caption{The message update order of parallel DS-BP$_4$ for the example in Fig.~\ref{fig:hatS}.
		(a) The initialization step, as well as the vertical step.
		(b) The horizontal step.
		The message update order will be iterated between (a) and (b).
	} \label{fig:PBP}
\end{figure}
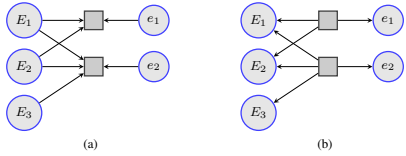
\begin{figure}[h!]	   % use \resizebox{rowsize}{colsize} and {!} means to maintain the aspect ratio
	\centering
	\resizebox{0.5\textwidth}{!}{ 
		\subfloat[\label{fig:SBP_1_1}]{\input{Fig_SBP_1_1.tex}}\qquad
		\subfloat[\label{fig:SBP_1_2}]{\input{Fig_SBP_1_2.tex}}\qquad\qquad
		\subfloat[\label{fig:SBP_2_1}]{\input{Fig_SBP_2_1.tex}}\qquad
		\subfloat[\label{fig:SBP_2_2}]{\input{Fig_SBP_2_2.tex}} 
	}
	\caption{The message update order of serial DS-BP$_4$ (along check nodes) for the example in Fig.~\ref{fig:hatS}.
		(a) and (b): serial update for check node 1.
		(c) and (d): serial update for check node 2.
		The message update order will be iterated from (a) to (d).
	} \label{fig:SBP}
\end{figure}
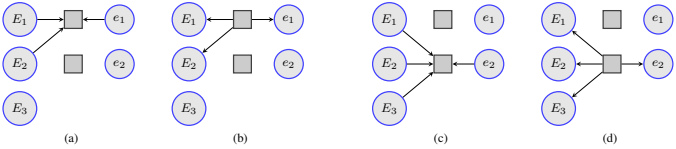

\subsection{Simulation Results}
We construct a $[[129,28]]$ CSS-type HP code \cite{TZ14} based on the $[7,4,3]$ and $[15,7,5]$ BCH codes as in \cite{LP19,KL20}.
This code has minimum distance  $d=3$ and can be a candidate for Theorem~\ref{thm:c2d3}.
Its raw check matrix has some columns of weight~one. After row multiplications by adjacent rows, we can obtain
   a check matrix, each column of which is of weight at least 2.\footnote{ Since the raw check matrix has a cyclic-like structure,   the multiplication  of two adjacent rows will not have high weight and the locality is   slightly affected.}  
   Then by Theorem~\ref{thm:c2d3}, we have a $[[129,28\,|\,101]]$ quantum HP DS code with  minimum distance $\tilde{d}=3$.

We first explain the serial schedule that will be conducted in the following simulations. 
In \cite{KL20}, it is demonstrated that  based on the raw check matrix of the $[[129,28]]$  HP code, the parallel BP$_4$ decoding does not perform well due to decoding oscillation (which is caused by the numerous short cycles and symmetric sub-graphs in the  Tanner graph \cite{KL20,RV20}); on the other hand,  the serial BP$_4$  along variable nodes performs quite well by using the raw matrix \cite{KL20}. 
We have created a check matrix so that  each of its column  has weight $\ge 2$ for Theorem~\ref{thm:c2d3}; however,  the serial update along variable nodes is too aggressive at certain variable node for the new check matrix (when computing the hard-decision and outgoing messages). This causes for some weight-one errors to be decoded as weight-three errors, and the syndrome is falsely matched.
Fortunately, this can be improved by using a serial update along the check nodes 
(which also breaks the symmetry in the short cycles and sub-graphs; however, it provides a more gradual update for each coordinate $n$ at each iteration). 

In this subsection, the simulation of BP allows a maximum number of 12 iterations, and each data point is based on collecting at least 100 blocks of logical errors.

We begin with the simulation without syndrome errors ($\epsilon_S=0$) and compare the decoding results of \mbox{parallel DS-BP$_4$} (Algorithm~\ref{alg:QBP}) and serial DS-BP$_4$ (Algorithm~\ref{alg:SBP}).
In this case,  DS-BP$_4$ is equivalent to the usual BP$_4$ in \cite{KL20}. 

We use bounded-distance decoding (BDD) as a benchmark for error performance.
In general, BDD with radius $t$ can correct any error of weight no larger than $t$.
We consider a more general BDD as follows.
Let $t\ge 0$ and $\gamma = (\gamma_0,\gamma_1,\dots,\gamma_t)$,
where $\gamma_j$ denotes the percentage of weight-$j$ errors assumed to be corrected.
Then the generalized BDD with respect to $t$ and $\gamma$ has a logical error rate at  $\epsilon$ ($=\epsilon_D$ here)  as follows 
\begin{equation} \label{eq:BDD}
P_\text{e,BDD}(N,t,\gamma) = 1- \textstyle \left(\sum_{j=0}^t \gamma_j\binom{N}{j} \epsilon^j (1-\epsilon)^{(N-j)} \right).
\end{equation}
The $[[129,28]]$ code  can correct any error of weight one.
If we make a lookup table for decoding by assigning each syndrome to low-weight errors, then  this code can 
correct about $98.73\%$ of the weight-2 errors.
Thus, with $t=2$, this lookup-table decoding provides a generalized BDD to have $\gamma_0=1,~\gamma_1=1, \text{ and } \gamma_2\approx98.73\%.$
This code is not degenerate; its stabilizers have weight larger than 3. 
Consequently these three $\gamma_j$ values are fixed  whether the degeneracy is considered  or not,
and they dominate the error performance.

\begin{figure}
	\begin{center}
		\includegraphics[scale = 0.6]{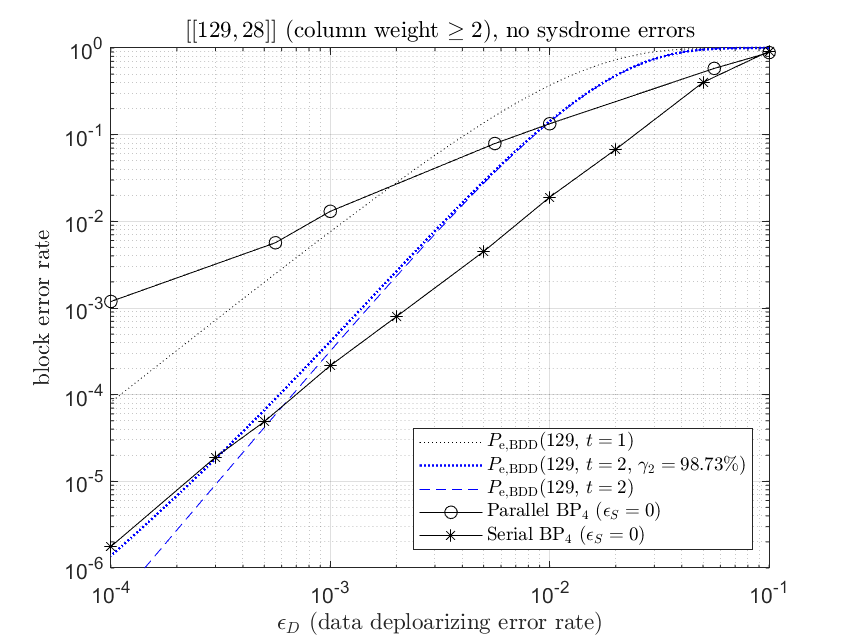}
		\caption{
			Decoding performance of the $[[129,28]]$ HP code (with syndrome error rate $\epsilon_S=0$). 
			The serial schedule is along the check nodes.
			Specific BDD reference performance curves, per \eqref{eq:BDD}, are plotted.
		}
		\label{fig:eps_S=0}
	\end{center}
\end{figure}

Figure~\ref{fig:eps_S=0} shows the simulations of   Algorithm~\ref{alg:QBP}  and  Algorithm~\ref{alg:SBP} at $\epsilon_S=0$,
together with   several BDD reference curves.  			 Note that $\gamma_j = 1$ for $j\leq t$ if not specified.
\mbox{Serial DS-BP$_4$} achieves a performance quite close to the lookup-table decoder $P_\text{e,BDD}(N,2,\gamma_2=98.73\%)$.
This matches Gallager's expectation that the performance of BP can be as close as to two times of the BDD performance with radius $t=\frac{d-1}{2}$ \cite{Gal63}.

\begin{figure}
	\begin{center}
		\includegraphics[scale = 0.6]{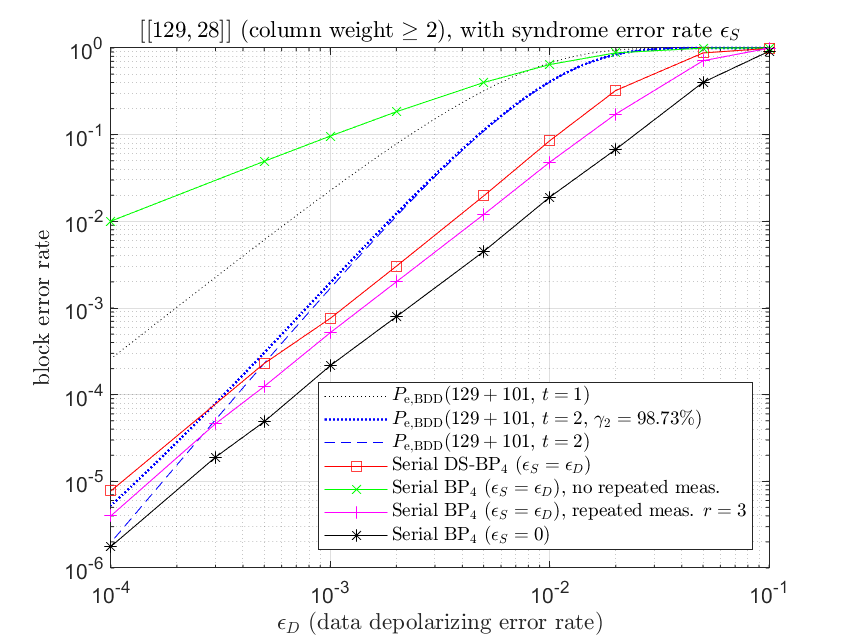}
		\caption{
			Decoding performance of the $[[129,28\,|\,101]]$ HP DS code. 
			All the serial-schedule results are based on Algorithm~\ref{alg:SBP}.
			When it is labeled with ``Serial BP$_4$'', it means that Algorithm~\ref{alg:SBP} is run with $\epsilon_S=0$ regardless of the actual $\epsilon_S$ value.
			If there are $r$ repeated measurements (meas.), a majority vote will be run to decide the syndrome before running the decoding algorithm.
		}
		\label{fig:DS-BP_eps}
	\end{center}
\end{figure}

Next, we assume that each syndrome bit is flipped with rate $\epsilon_S\neq 0$.
For simplicity, assume $\epsilon_S = \epsilon_D$, which allows us to use \eqref{eq:BDD} as a benchmark.
We focus on the serial schedule since it provides a better performance.
The serial DS-BP$_4$ performance  is plotted in Fig.~\ref{fig:DS-BP_eps},
which has a performance loss of less than an order compared to the case of no syndrome error  $\epsilon_S=0$.
It can be seen that serial DS-BP$_4$ improves  serial BP$_4$ ($\epsilon_S = \epsilon_D$) if no repeated measurements are conducted as expected.
We also provide a curve for serial BP$_4$ ($\epsilon_S = \epsilon_D$) with $r=3$ repeated measurements, and it performs quite well as seen in Fig.~\ref{fig:DS-BP_eps};
however, this only reveals the importance of eliminating the effect caused by noisy measurements.

\begin{figure}
	\begin{center}
		\includegraphics[scale = 0.6]{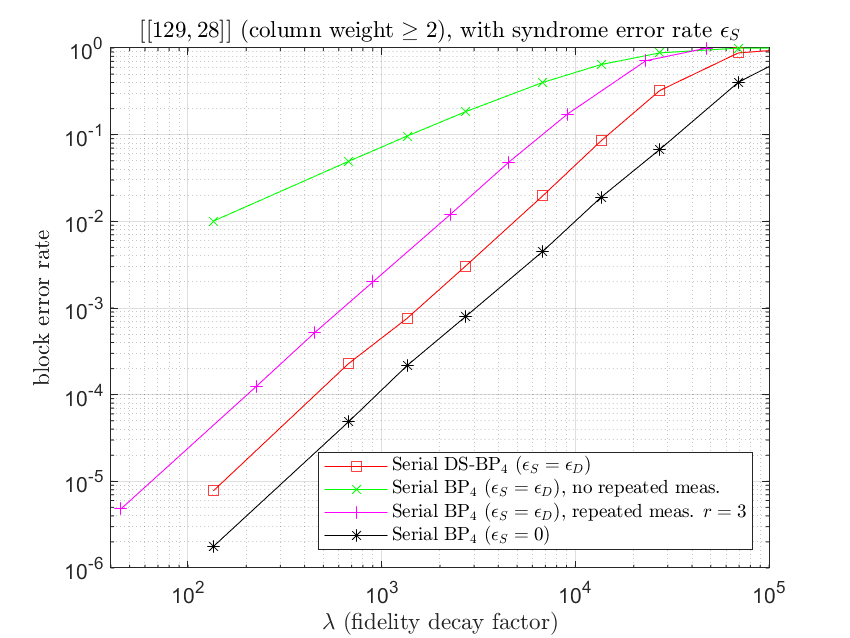}
		\caption{
			Comparing the different decoding strategies in Fig.~\ref{fig:DS-BP_eps}, at a certain fidelity decay factor $\lambda$ as in \eqref{eq:loss}, in which we assume that $\epsilon = \epsilon_D$ and $\tau=r\times 740$~ns (where there is only one case with $r=3$, as labeled, and the other cases have $r=1$).
		}
		\label{fig:DS-BP_loss}
	\end{center}
\end{figure}

Performing repeated measurements would require additional time (and gates), so these practical issues should be considered in comparison.
As described in \cite{MZLL18,Nic+17}, the fidelity of a physical qubit   decays exponentially over  the operational time~$\tau$.
Assume that the fidelity is 
\begin{equation} \label{eq:loss}
1 - \epsilon = e^{-\lambda\tau}
\end{equation}
for some decay factor $\lambda$.
Suppose that a round of syndrome measurement takes 740~ns \cite{Ver+17}.
We further assume that the measurement time dominates the overall error-correction time, since the decoder should run much faster in a classical hardware~\cite{VCB17}.
So now one round of measurements with one round of serial DS-BP$_4$ takes about $740$~ns, and   $r=3$ rounds of measurements with one round of serial BP$_4$ takes about $3\times 740$~ns.
Given $\epsilon$ ($= \epsilon_D$ in Fig.~\ref{fig:DS-BP_eps}) and $\tau$ ($=740$~or~$3\times 740$~ns), a corresponding $\lambda$ in \eqref{eq:loss} can be derived.
 Figure~\ref{fig:DS-BP_loss} provides the rescaled curves of Fig.~\ref{fig:DS-BP_eps}. 
The results show that using the DS-BP approach can take the advantage of less measurement time to outperform a decoding strategy with repeated measurements.

\section{Conclusion \& Future Works} \label{sec:Conc}

Faulty syndrome measurement is an issue that cannot be neglected in fault-tolerant quantum error-correction.  A potential solution is to apply quantum DS codes.
By generalizing the refined BP$_4$ \cite{KL20}, we proposed a DS-BP$_4$ decoding algorithm and have demonstrated that it can efficiently  achieve satisfactory results.
The DS-BP approach has proved to be suitable for low-weight stabilizers and can be used with less (or even without) redundant measurements. 
This decreases the measurement time and increases the probability of a successful decoding.

We simulated the $[[129,28 \,|\,101]]$ HP DS code with minimum distance 3.
For codes with higher minimum distance, we may use additional syndrome measurements to compensate the effects of syndrome errors.
For example, the surface codes are the current state-of-the-art candidate for FTQC.
It is still unknown whether BP works for codes with strong degeneracy that their stabilizers may have weight much lower than the minimum distance.
Also, the error model considered in this paper is too ideal. 
To apply the DS-BP approach in a more practical model, like the faulty circuit model~\cite{Fow+12,ZAWP19}, 
is our ongoing research.

\section*{Acknowledgment}
%We are indebted to Michael Shell for maintaining and improving
%\texttt{IEEEtran.cls}. 

CYL was financially supported from the Young Scholar Fellowship Program by the Ministry of Science and Technology (MOST) in Taiwan, under Grant MOST109-2636-E-009-004.

%%%%%%
%% To balance the columns at the last page of the paper use this
%% command:
%%
%\enlargethispage{-1.2cm} 
%%
%% If the balancing should occur in the middle of the references, use
%% the following trigger:
%%
%%\IEEEtriggeratref{3}
%%
%% which triggers a \newpage (i.e., new column) just before the given
%% reference number. Note that you need to adapt this if you modify
%% the paper.  The "triggered" command can be changed if desired:
%%
%\IEEEtriggercmd{\enlargethispage{-20cm}}
%%
%%%%%%

%%%%%%
%% References:
%% We recommend the usage of BibTeX:
%%
%\bibliographystyle{IEEEtran}
%\bibliography{definitions,bibliofile}
%%
%% where we here have assume the existence of the files
%% definitions.bib and bibliofile.bib.
%% BibTeX documentation can be obtained at:
%% http://www.ctan.org/tex-archive/biblio/bibtex/contrib/doc/
%%%%%%

\bibliographystyle{ieeetr}  % 使用 IEEE Trans 期刊格式
\bibliography{bibb}  %reference 所需的bib檔

%% Or you use manual references (pay attention to consistency and the
%% formatting style!):

\end{document}

%% file: Fig_3bits_DS.tex
\begin{tikzpicture}[node distance=1.3cm,>=stealth,bend angle=45,auto]

\tikzstyle{chk}=[rectangle,thick,draw=black!75,fill=black!20,minimum size=4mm]
\tikzstyle{var}=[circle,thick,draw=blue!75,fill=gray!20,minimum size=4mm,font=\footnotesize]
\tikzstyle{VAR}=[circle,thick,draw=blue!75,fill=blue!20,minimum size=5mm,font=\footnotesize]
\tikzstyle{fac}=[anchor=west,font=\footnotesize]

\node[var] (x3) at (0,0) {$E_3$};
\node[var] (x2) at (0,1) {$E_2$};
\node[var] (x1) at (0,2) {$E_1$};
\node[chk] (c1) at (1.5,2) {};
\node[chk] (c2) at (1.5,1) {};
\node[var] (z1) at (2.8,2) {$e_1$};
\node[var] (z2) at (2.8,1) {$e_2$};

\draw[thick] (x1) -- (c1);
\draw[thick,dashed] (x2) -- (c1);
\draw[thick,densely dotted] (x1) -- (c2) -- (x2);
\draw[thick,dashed] (x3) -- (c2);
\draw[thin] (c1) -- (z1);
\draw[thin] (c2) -- (z2);

\node[fac] [right of=c1,xshift=9mm,yshift=6.5mm] {$(\tilde S_1):~\langle E_1,X\rangle_{\sG}+\langle E_2,Y\rangle_{\sG} + e_1= z_1$};
\node[fac] [right of=c2,xshift=16mm,yshift=-6.7mm] {$(\tilde S_2):~\langle E_1,Z\rangle_{\sG}+\langle E_2,Z\rangle_{\sG}+\langle E_3,Y\rangle_{\sG} + e_2 = z_2$};

\node[fac] (Xl) [right of=z2,xshift=5mm, yshift= 7+14] {};
\node[fac] (Xr) [right of=z2,xshift=20mm,yshift= 7+14] {$X$};
\draw[thick] (Xl) -- (Xr);
\node[fac] (Yl) [right of=z2,xshift=5mm, yshift= 0+14] {};
\node[fac] (Yr) [right of=z2,xshift=20mm,yshift= 0+14] {$Y$};
\draw[thick,dashed] (Yl) -- (Yr);
\node[fac] (Zl) [right of=z2,xshift=5mm, yshift=-7+14] {};
\node[fac] (Zr) [right of=z2,xshift=20mm,yshift=-7+14] {$Z$};
\draw[thick,densely dotted] (Zl) -- (Zr);

\end{tikzpicture}

%% file: Fig_PBP_1.tex
\begin{tikzpicture}[node distance=1.3cm,>=stealth,bend angle=45,auto]

\tikzstyle{chk}=[rectangle,thick,draw=black!75,fill=black!20,minimum size=4mm]
\tikzstyle{var}=[circle,thick,draw=blue!75,fill=gray!20,minimum size=4mm,font=\footnotesize]
\tikzstyle{VAR}=[circle,thick,draw=blue!75,fill=blue!20,minimum size=5mm,font=\footnotesize]
\tikzstyle{fac}=[anchor=west,font=\footnotesize]

\node[var] (x3) at (0,0) {$E_3$};
\node[var] (x2) at (0,1) {$E_2$};
\node[var] (x1) at (0,2) {$E_1$};
\node[chk] (c1) at (1.5,2) {};
\node[chk] (c2) at (1.5,1) {};
\node[var] (z1) at (2.8,2) {$e_1$};
\node[var] (z2) at (2.8,1) {$e_2$};

\draw[->] (x1) -- (c1);
\draw[->] (x2) -- (c1);
\draw[<-] (c1) -- (z1);
\draw[->] (x1) -- (c2);
\draw[->] (x2) -- (c2);
\draw[->] (x3) -- (c2);
\draw[<-] (c2) -- (z2);

\end{tikzpicture}

%% file: Fig_PBP_2.tex
\begin{tikzpicture}[node distance=1.3cm,>=stealth,bend angle=45,auto]

\tikzstyle{chk}=[rectangle,thick,draw=black!75,fill=black!20,minimum size=4mm]
\tikzstyle{var}=[circle,thick,draw=blue!75,fill=gray!20,minimum size=4mm,font=\footnotesize]
\tikzstyle{VAR}=[circle,thick,draw=blue!75,fill=blue!20,minimum size=5mm,font=\footnotesize]
\tikzstyle{fac}=[anchor=west,font=\footnotesize]

\node[var] (x3) at (0,0) {$E_3$};
\node[var] (x2) at (0,1) {$E_2$};
\node[var] (x1) at (0,2) {$E_1$};
\node[chk] (c1) at (1.5,2) {};
\node[chk] (c2) at (1.5,1) {};
\node[var] (z1) at (2.8,2) {$e_1$};
\node[var] (z2) at (2.8,1) {$e_2$};

\draw[<-] (x1) -- (c1);
\draw[<-] (x2) -- (c1);
\draw[->] (c1) -- (z1);
\draw[<-] (x1) -- (c2);
\draw[<-] (x2) -- (c2);
\draw[<-] (x3) -- (c2);
\draw[->] (c2) -- (z2);

\end{tikzpicture}

%% file: Fig_SBP_1_1.tex
\begin{tikzpicture}[node distance=1.3cm,>=stealth,bend angle=45,auto]

\tikzstyle{chk}=[rectangle,thick,draw=black!75,fill=black!20,minimum size=4mm]
\tikzstyle{var}=[circle,thick,draw=blue!75,fill=gray!20,minimum size=4mm,font=\footnotesize]
\tikzstyle{VAR}=[circle,thick,draw=blue!75,fill=blue!20,minimum size=5mm,font=\footnotesize]
\tikzstyle{fac}=[anchor=west,font=\footnotesize]

\node[var] (x3) at (0,0) {$E_3$};
\node[var] (x2) at (0,1) {$E_2$};
\node[var] (x1) at (0,2) {$E_1$};
\node[chk] (c1) at (1.5*4/5,2) {};
\node[chk] (c2) at (1.5*4/5,1) {};
\node[var] (z1) at (2.8*4/5,2) {$e_1$};
\node[var] (z2) at (2.8*4/5,1) {$e_2$};

\draw[->] (x1) -- (c1);
\draw[->] (x2) -- (c1);
\draw[<-] (c1) -- (z1);
%\draw[->] (x1) -- (c2);
%\draw[->] (x2) -- (c2);
%\draw[->] (x3) -- (c2);
%\draw[<-] (c2) -- (z2);

\end{tikzpicture}

%% file: Fig_SBP_1_2.tex
\begin{tikzpicture}[node distance=1.3cm,>=stealth,bend angle=45,auto]

\tikzstyle{chk}=[rectangle,thick,draw=black!75,fill=black!20,minimum size=4mm]
\tikzstyle{var}=[circle,thick,draw=blue!75,fill=gray!20,minimum size=4mm,font=\footnotesize]
\tikzstyle{VAR}=[circle,thick,draw=blue!75,fill=blue!20,minimum size=5mm,font=\footnotesize]
\tikzstyle{fac}=[anchor=west,font=\footnotesize]

\node[var] (x3) at (0,0) {$E_3$};
\node[var] (x2) at (0,1) {$E_2$};
\node[var] (x1) at (0,2) {$E_1$};
\node[chk] (c1) at (1.5*4/5,2) {};
\node[chk] (c2) at (1.5*4/5,1) {};
\node[var] (z1) at (2.8*4/5,2) {$e_1$};
\node[var] (z2) at (2.8*4/5,1) {$e_2$};

\draw[<-] (x1) -- (c1);
\draw[<-] (x2) -- (c1);
\draw[->] (c1) -- (z1);
%\draw[->] (x1) -- (c2);
%\draw[->] (x2) -- (c2);
%\draw[->] (x3) -- (c2);
%\draw[<-] (c2) -- (z2);

\end{tikzpicture}

%% file: Fig_SBP_2_1.tex
\begin{tikzpicture}[node distance=1.3cm,>=stealth,bend angle=45,auto]

\tikzstyle{chk}=[rectangle,thick,draw=black!75,fill=black!20,minimum size=4mm]
\tikzstyle{var}=[circle,thick,draw=blue!75,fill=gray!20,minimum size=4mm,font=\footnotesize]
\tikzstyle{VAR}=[circle,thick,draw=blue!75,fill=blue!20,minimum size=5mm,font=\footnotesize]
\tikzstyle{fac}=[anchor=west,font=\footnotesize]

\node[var] (x3) at (0,0) {$E_3$};
\node[var] (x2) at (0,1) {$E_2$};
\node[var] (x1) at (0,2) {$E_1$};
\node[chk] (c1) at (1.5*4/5,2) {};
\node[chk] (c2) at (1.5*4/5,1) {};
\node[var] (z1) at (2.8*4/5,2) {$e_1$};
\node[var] (z2) at (2.8*4/5,1) {$e_2$};

%\draw[->] (x1) -- (c1);
%\draw[->] (x2) -- (c1);
%\draw[<-] (c1) -- (z1);
\draw[->] (x1) -- (c2);
\draw[->] (x2) -- (c2);
\draw[->] (x3) -- (c2);
\draw[<-] (c2) -- (z2);

\end{tikzpicture}

%% file: Fig_SBP_2_2.tex
\begin{tikzpicture}[node distance=1.3cm,>=stealth,bend angle=45,auto]

\tikzstyle{chk}=[rectangle,thick,draw=black!75,fill=black!20,minimum size=4mm]
\tikzstyle{var}=[circle,thick,draw=blue!75,fill=gray!20,minimum size=4mm,font=\footnotesize]
\tikzstyle{VAR}=[circle,thick,draw=blue!75,fill=blue!20,minimum size=5mm,font=\footnotesize]
\tikzstyle{fac}=[anchor=west,font=\footnotesize]

\node[var] (x3) at (0,0) {$E_3$};
\node[var] (x2) at (0,1) {$E_2$};
\node[var] (x1) at (0,2) {$E_1$};
\node[chk] (c1) at (1.5*4/5,2) {};
\node[chk] (c2) at (1.5*4/5,1) {};
\node[var] (z1) at (2.8*4/5,2) {$e_1$};
\node[var] (z2) at (2.8*4/5,1) {$e_2$};

%\draw[->] (x1) -- (c1);
%\draw[->] (x2) -- (c1);
%\draw[<-] (c1) -- (z1);
\draw[<-] (x1) -- (c2);
\draw[<-] (x2) -- (c2);
\draw[<-] (x3) -- (c2);
\draw[->] (c2) -- (z2);

\end{tikzpicture}